\documentclass[12pt,a4paper]{article}

\newcommand{\lp}{\left}
\newcommand{\rp}{\right}
\newcommand{\be}{\begin{equation}}
\newcommand{\ee}{\end{equation}}

\newcommand{\eeqa}{\end{eqnarray}}

\newcommand{\dep}{\partial}
\newcommand{\al}{\alpha}
\newcommand{\ch}{\chi}
\newcommand{\sg}{\sigma}
\newcommand{\tht}{\theta}
\newcommand{\ovr}{\overline}
\newcommand{\sgb}{\overline{\sigma}}
\newcommand{\chb}{\overline{\chi}}
\newcommand{\phb}{\overline{\phi}}
\newcommand{\rhb}{\overline{\rho}}
\newcommand{\cb}{\overline{c}}

\begin{document}

\title{Noncompact gauge fields on a lattice: SU(n) theories}
\author{
  { Fabrizio Palumbo~\thanks{This work has been partially
  supported by EEC under RTN contract HPRN-CT-2000-00131}}        \\
  {\small\it INFN -- Laboratori Nazionali di Frascati}  \\[-0.2cm]
  {\small\it P.~O.~Box 13, I-00044 Frascati, ITALIA}          \\[-0.2cm]
  {\small Internet: {\tt palumbof@lnf.infn.it}}
  \\[-0.2cm]
  \\[-0.1cm]  \and
  { Roberto Scimia }             \\
  {\small\it Marconi Mobile - ELMER S.p.A.} \\[-0.2cm]
  {\small\it Viale dell'Industria, 4 - 00040 Pomezia (Roma), ITALIA}
\\[-0.2cm]
  {\small\it Dipartimento di Fisica and INFN -- Sezione di Perugia}
\\[-0.2cm]
  {\small\it Universit\`a degli Studi di Perugia}        \\[-0.2cm]
  {\small\it Via A. Pascoli - 06100 Perugia, ITALIA}          \\[-0.2cm]
  {\small Internet: {\tt roberto.scimia@lnf.infn.it}}
   }

\maketitle

\thispagestyle{empty}   

\begin{abstract}
Recently it has been found that in a noncompact lattice
regularization of the SU(2) gauge theory the physical volume is
larger than in the Wilson theory with the same 
number of sites.
In its original formulation  the noncompact regularization is directly
applicable to U(n) theories for any n and to SU(n) theories for n=2.
In this paper we extend it to SU(n) for any n and
investigate some of its properties.
\end{abstract}

\clearpage

\section{Introduction}

One of the present problems of lattice gauge theories is how to
increase the physical volume where the numerical simulations
are performed. The physical size of the lattice is indeed
a major limitation in the study of hadronic structure
functions~\cite{volume} and light hadron spectroscopy~\cite{spectro},
and in the evaluation of the ratio $\epsilon' / \epsilon$~\cite{Lell}.

A recent investigation showed that for SU(2)
lattice gauge theory  a noncompact regularization provides a
physical volume larger than the Wilson theory with the same
number of sites~\cite{GDC}. It appears therefore interesting to
repeat the simulation for the physically relevant SU(3) theory, but
in its original  formulation  this regularization  is directly
applicable to U(n) theories for any n but to SU(n) theories  for n=2 only. It is
the purpose of the present paper to extend it to SU(n) for any n and
to investigate some of its properties.

As it is well known a formal discretization of gauge theories
breaks gauge invariance. To avoid this inconvenient Wilson
assumed~\cite{Wil} as
dynamical variables elements of the gauge group instead of the gauge
fields which live in the group algebra. In this way one gets a theory with an
exact symmetry which has the desired formal continuum limit. This theory is said
compact because compact are the dynamical variables.

The success of Wilson's regularization is by now celebrated
in textbooks.
But one can wonder whether its exact lattice symmetry can also
be realized
without compactifying the variables, and if this can have some
advantages
wrt specific issues, reducing the artifacts of the lattice.
In addition to the mentioned possibility of having larger
volumes, the importance of noncompact gauge fields, especially in
their coupling with matter fields, has been advocated in the
investigation of a possible fixed point of QED
at a finite coupling~\cite{kogut}. Moreover
perturbative calculations should be easier since one does not
have to expand the link variables of the Wilson theory
in terms of the gauge fields. Perturbative calculations are at least necessary
to make contact  with the continuum  formulation, but other applications
like the study of renormalons should also be mentioned~\cite{marc}.
Finally in numerical simulations
one might expect a faster approach to the scaling, the more
so the more important is the summation of the tadpoles~\cite{Lep}
generated by the expansion of the link variables.

If one defines the covariant  derivative in close analogy
to the continuum
as an ordinary discrete  derivative plus the appropriate
element of the algebra
of the group,  the lattice symmetry is broken, but it can
be maintained by
introducing  compensating auxiliary fields which decouple
in the continuum limit~\cite{Fab,BeFab1}.
Such a regularization has been
studied exhaustively in
the case of SU(2). Specifically the renormalization group
parameter has been
evaluated, and the perturbative properties have been shown
to agree with
Lusher's calculation in Wilson regularization~\cite{FBnp,BeFab2}. Moreover
Monte Carlo simulations
gave results compatible with Wilson's theory, but
with the interesting difference mentioned at the beginning
: the physical volume  results larger than in
the Wilson theory  with
the same number of sites~\cite{GDC}. This is what we expect
euristically for a regularization closer to the continuum.

In the original formulation of this noncompact regularization,
with the exception of the case n=2,
invariance wrt SU(n) implies invariance wrt U(n). The aim of
this paper is to construct a potential which breaks the U(n)
invariance down to SU(n) and to investigate some of its properties.
 For simplicity, explicit formulae will be given for
n=3, but the generalization is obvious. Obvious is also the
coupling to matter fields wich will therefore not be discussed.

 In Section 2 we report, for the convenience of the reader,
the regularization
for U(n).  In Section 3 we show
how to construct a
SU(n) invariant theory. In Sections 4 and 5 we derive the Ward
identities and
the formulation in a background gauge, which can be useful in
perturbative
calculations. In the Appendix we report the explicit expression
of the U(3) breaking potential.

\section{The  noncompact regularization for U(3)}

We first consider the regularization of U(n) gauge theories. For
n=1 we get a truly noncompact QED, namely noncompact also in the
coupling with the matter fields.

We want to construct a covariant derivative ${\cal {D}}_{\mu}$ which
transforms according to
\be
{\cal {D}}_\mu \, ' (x)  = g(x) {\cal {D}}_\mu (x)
g^{\dagger} (x+\mu) \label{tradeco}
\ee
when $g(x)$ is an element of U(n). A simple discretization of the
continuum would give
\be
{\cal {D}}_\mu = \nabla_{\mu} + i \left( \ch_{\mu}1\!\!1
+ A^a_{\mu}T_a \right),
\ee
where $\nabla_{\mu} $ is the ordinary right discrete derivative, and
$\chi_{\mu}$ and $A^a_{\mu}$ are the abelian and nonabelian gauge fields
respectively. We adopt for the generators of $SU(3)$ in the fundamental
representation  the normalizations\footnote{These normalizations are
slightly
different from those used in~\cite{BeFab1}.}
\be
\lp \{T_a\, , T_b \rp \} = \frac{4}{3} \delta_{ab}\,1\!\!1
+ 2 d^c_{ab}\,T_c\, ,
\;\;\;\; \lp [T_a\, , T_b \rp ] = 2if^c_{ab}\, T_c .
\ee

As it is well known with such a definition of the covariant
derivative it is impossible to fulfill the transformation
rule of Eq.~(\ref{tradeco}). The way out that we reconsider here is
based on the use
of auxiliary compensating fields. It turns out that the lattice
covariant derivative transforms in the right way if it acts
on a field $\psi$ in the fundamental representation according to
\be
\left({\cal {D}}_\mu  \psi \right) (x) = D_{\mu}(x) \psi(x+\mu) -
{ 1\over a}  \psi(x),
\ee
where $D_{\mu}$ has the following form
\be
D_\mu(x) = \left[\frac{1}{a} -\sg_\mu (x) +i\ch_\mu (x) \right] 1\!\!1
+ \left[ i A^a_\mu  -\al^a_\mu(x) \right] T_a .
\ee
In the above equation $"a"$ is the lattice spacing, and $\sg_{\mu}\;$
and $\al_{\mu}\;$ are the additional fields necessary to enforce the
lattice gauge invariance . With a little abuse of language we
will call also $D_{\mu}$ covariant derivative. The action of U(3) on the
fields, for $g(x)\simeq 1\!\!1-iT_a\tht^a(x)-i1\!\!1\tht^0(x)$, is
\begin{eqnarray} \lp (A^a_\mu (x) \rp)' & = & A^a_\mu (x) +\Delta_\mu \tht
^a(x)
+2f^a_{bc}\tht ^b(x)A^c_\mu (x)  -a\sg _\mu(x)\Delta_\mu \tht ^a(x)
\nonumber \\
  &   & -af^a_{bc}A^b_\mu (x)\Delta_\mu \tht ^c(x)
-a d^a_{bc}\al ^b_\mu(x)\Delta_\mu \tht ^c(x)
-a\al^a_\mu (x)\Delta_\mu \tht^0(x) \nonumber
\end{eqnarray}
\begin{eqnarray}
\lp (\al^a_\mu (x) \rp)' & = & \al^a_\mu (x)
+2f^a_{bc}\tht ^b(x)\al^c_\mu (x)
-af^a_{bc}\al^b_\mu (x)\Delta_\mu \tht ^c(x) \nonumber \\
  &   & +a\ch _\mu(x)\Delta_\mu \tht ^a(x)
+a d^a_{bc} A^b_\mu(x)\Delta_\mu \tht ^c(x)
+aA^a_\mu (x)\Delta_\mu \tht^0(x) \nonumber
\end{eqnarray}
\begin{eqnarray}
\lp (\ch_\mu (x) \rp)'  & = & \ch_\mu (x)
+\Delta_\mu \tht^0(x)
-\frac{2}{3}a\al^a_\mu (x)\Delta_\mu \tht ^a(x)
-a\sg_\mu (x)\Delta_\mu \tht^0(x) \nonumber \\
\lp (\sg_\mu (x) \rp)'  & = &  \sg_\mu (x)
+\frac{2}{3}aA^a_\mu (x)\Delta_\mu \tht ^a(x)
+a\ch_\mu (x)\Delta_\mu \tht^0(x) .\label{trans}
\end{eqnarray}
Since all the fields are mixed by the gauge transformations, we cannot say
at this point which are the physical fields. They are selected by the
action as we will see by studying the Ward identities.

A lattice action invariant under the above transformations is
\be
{\cal L}_{YM}(x)= { 1 \over 4} \beta\, \mbox{Tr} F_{\mu \nu}^+
F_{\mu \nu}, \label{YM}
\ee
where $ F_{\mu \nu}$ is the stress tensor
\be
F_{\mu \nu}(x) =  D_{\mu}(x) D_{\nu}(x+\mu) -
            D_{\nu}(x) D_{\mu}(x+\nu) .
\ee
We emphasize that in such a formulation the measure in the partition
function is flat.

 In the formal continuum limit the field $\sigma_{\mu}$
becomes invariant and decouples together with $\alpha_{\mu}$, so that
these seem to be the auxiliary fields. But the situation can be
different at the quantum level. To control the decoupling of the
redundant fields in the presence of quantum effects we use
the fact that in a noncompact regularization, besides ${\cal L}_{YM}$,
there are other local invariants, which can be used to construct an appropriate
potential and to give divergent masses to the fields which must
stay decoupled. One such potential is
\begin{eqnarray}
{\cal{L}}_1 & = & \beta_1 \sum_\mu Tr\,\lp[
D_\mu^{\dagger}(x)D_\mu (x)  -\frac{1\!\!1}{a^2}\rp]^2 \nonumber \\
  & = & \beta_1\sum_\mu \lp \{ \frac{12}{a^2}\sg^2_\mu(x)+
\frac{8}{a^2}\al^a_\mu(x)\al^a_\mu(x)
-\frac{12}{a} \sg_\mu(x)\lp ( \sg_\mu^2(x)+\ch_\mu^2(x)\rp )\rp .
\nonumber \\
  &   & -\frac{8}{a}\lp ( 3\sg_\mu(x)\al_\mu^a (x)\al_\mu^a (x)
+2\sg_\mu(x) A_\mu^a (x)A_\mu^a (x)
+2\ch_\mu (x) A_\mu^a (x)\al_\mu^a (x)   \rp )
\nonumber \\
  &   & +3\lp (\sg_\mu^2(x)+\ch_\mu^2(x) \rp )^2
+{4\over 3}\lp (A_\mu^2(x)+\al_\mu^2 (x) \rp )^2
+4A_\mu^a (x)A_\mu^a (x)\lp ( \sg_\mu^2 (x)+3\ch_\mu^2 (x)\rp )
\nonumber \\
  &   & +4\al_\mu^a (x)\al_\mu^a (x)\lp (3\sg_\mu^2 (x)
+\ch_\mu^2 (x)\rp )
+16\sg_\mu (x)\ch_\mu(x)A_\mu^a (x)\al_\mu^a (x)
\nonumber \\
  &   & +8d^a_{bc}\lp [\lp (A^a_\mu (x)A^b_\mu (x)
+\al^a_\mu (x)\al^b_\mu (x)\rp )
\lp(\sg_\mu (x)\al_\mu^c (x)+\ch_\mu(x)A_\mu^c (x)
-\frac{1}{a}\al^c_\mu (x)\rp) \rp ]
\nonumber \\
  &   &
+8f^h_{ab}f^h_{cd}A^a_\mu (x)\al^b_\mu (x)
A^c_\mu (x)\al^d_\mu (x)
+2d^h_{ab}d^h_{cd}\lp(A^a_\mu (x)A^b_\mu (x)
A^c_\mu (x)A^d_\mu (x)\rp .
\nonumber \\
  &   & \lp . +\al^a_\mu (x)\al^b_\mu (x)
\al^c_\mu (x)\al^d_\mu (x)
+2A^a_\mu (x)A^b_\mu (x)
\al^c_\mu (x)\al^d_\mu (x)\rp )
\nonumber \\
  &   &  \lp . +8d^h_{ab}f^h_{cd}A^a_\mu (x)
\al^b_\mu (x)\lp ( A^c_\mu (x)A^d_\mu (x)+\al^c_\mu (x)\al^d_\mu (x)
\rp )
\rp \}.
\end{eqnarray}
We see that ${\cal{L}}_1 $ provides the desired divergent masses to
the auxiliary fields in the trivial vacuum. A more general analysis
of the mass spectrum will be given in Section 4.
There are other invariant terms, which can be used for instance to make
the propagator of some of the auxiliary fields strictly local on the lattice
\cite{FBnp}, but we will ignore them for simplicity.

The effect of ${\cal{L}}_1$ can be well understood
by adopting a definition of the covariant derivative where the
abelian fields are in a polar representation
\be
D_\mu(x) = \hat{D}_{\mu}(x) \exp\,i \phi_{\mu}(x), \label{polar}
\ee
where
\be
\hat{D}_{\mu} = \rho_{\mu} 1\!\!1 + \left[ i( A')^a_\mu
-(\al ')^a_\mu \right] T_a .
\ee
Due to ${\cal{L}}_1$, the $\rho$-field acquires a non vanishing expectation
value $<\rho_{\mu}>= 1/a$. The U(3) symmetry is "spontaneously" broken,
and the components of $\phi_{\mu}$ are the Goldston bosons\footnote
{Needless to say, the U(3) symmetry remains exact. While for
$<\rho_{\mu}>=0$ it is realized linearly, for $<\rho_{\mu}>\neq 0$ it is
realized nonlinearly.}.
As we will see by studying the Ward identities, the physical fields
are~$\phi_{\mu}$ and~$A'_\mu$.

It is worth while noticing that in the absence of the spontaneous
symmetry breaking there is not even a discrete derivative, the term 1/a 
being absent in the definition of $D_{\mu}$. The present
definition of gauge theories on a lattice can then be regarded as
a matrix model where the space-time dynamics is generated by a spontaneous
 breaking of the gauge symmetry.

\section{The noncompact regularization for SU(n)}

A derivative covariant wrt SU(n) transformations only, must in
general contain all the fields of the U(n) theory, the only
difference being that the field
$\ch_{\mu}$ becomes another auxiliary field. So we cannot restrict
ourselves to the SU(n) symmetry by changing the covariant derivative, and at the
same time the potential ${\cal{L}}_1$ does not generate a mass for
the $\chi$-field. Moreover, as it will be confirmed in
the next Section by the Ward identities, no U(n) invariant
potential can generate a mass for both abelian
fields. We must therefore break explicitely the U(3) symmetry
in order to give to the would be Goldstone bosons a mass,
actually a divergent mass.

The case n=2 is exceptional, because for SU(2)
transformations, namely for $\theta_0 = 0$, Eqs.~\ref{trans}
do not mix the multiplet  $A_{\mu}, \sigma_{\mu}$ with the
multiplet $\al_{\mu}, \chi_{\mu}$. Therefore we can break
U(2) by omitting the latter fields to get an SU(2) invariant
theory. This case has already been exhaustively studied
\cite{BeFab1,FBnp,BeFab2}.

There are two terms ( whose expression wil be spelled out in the
Appendix) which break the U(3) invariance of the action, explicitely
\begin{eqnarray}
{\cal{L}}_2 & = & \beta_2 { 1\over a} \sum_\mu
\lp [  \mbox{det} \,D_\mu(x) + \mbox{det} \,D_\mu^{\dagger}(x) \rp]
\\
{\cal{L}}_2\!' & = & \beta_2 \!' { i\over a} \sum_\mu
\lp [ \mbox{det} \,D_\mu(x) - \mbox{det} \,D_\mu^{\dagger}(x) \rp] .
\end{eqnarray}
 But we can always
get rid of one of them by the global trasformation
\be
D_{\mu}= D'_{\mu} \exp\, i \alpha_{\mu}.
\ee
For instance, we can get rid of ${\cal{L}}_2\!' $ by setting in the
above equation $ \alpha= 1/3 \;\mbox{arctg}\;(\beta_2 / \beta_2')$.
We assume this to be the case.

We now determine the minima of the action at constant fields in the presence of
${\cal{L}}_2$. We assume that the color symmetry is not spontaneously broken. As
a consequence the colored fields cannot develop a nonvanishing expectation
value, neither can they mix with the auxiliary abelian fields. By adopting
the abelian polar representation of Eq.~(\ref{polar}) we minimize ${\cal{L}}_2$
wrt $\phb_\mu$ at
fixed $\rhb_\mu$, and then minimize the resulting action wrt $\rhb_\mu$.

By noticing that
\be
\ovr{\cal{L}}_2=\beta_2\;\frac{2}{a}\;\rhb_\mu^3 \;
\cos\lp (3\phb_\mu\rp )
\ee
we obtain the stationarity condition
\be
 \sin 3\phb_\mu=0.
\ee
Assuming $\beta_2 < 0$, the minimum of ${\cal{L}}_2$  occurs at
$\phb_{\mu} = 0, 2\pi/3,4\pi/3$, namely
the covariant derivative at the minimum belongs to
the center of SU(3)\footnote{All the minima are therefore in one-to-one
correspondence with those of the Wilson theory, and the difficulty
raised in ref.~\cite{FBnp} in connection with this degeneracy can then be
overcome as in the compact regularization.}.

Nex we require, as a normalization condition, that the total
action have one and only one minimum at $\rhb_\mu = 1$.
To achieve this result we find it necessary to add another
potential term
\begin{eqnarray}
{\cal{L}}_3 & = & \beta_3 \frac{1}{a^2} \sum_\mu Tr\,\lp[
D_\mu^{\dagger}(x)D_\mu (x)  -\frac{1\!\!1}{a^2}\rp] \nonumber \\
& = & \beta_3 { 1 \over a^2}\sum_\mu \lp [ -\frac{6}{a} \sg_\mu (x)
+3\sg^2_\mu(x)+3\ch^2_\mu(x) \rp .
\nonumber \\
  &   & \lp . +2\al^a_\mu(x)\al^a_\mu(x)
+2A^a_\mu(x)A^a_\mu(x) \rp ] .
\end{eqnarray}
This term seems to give a mass also to all the colored fields,
but it has
already been shown that this is not the case for SU(2), and the  proof
will be generalized in the next Section.

Taking into account that at the minimum
\be
\ovr{\cal{L}}_2 = -2 \frac{|\beta_2|}{a} \, \overline{\rho}_{\mu}^3,
\ee
we then have, omitting some constant terms
\be
\ovr{\cal{L}}=\sum_\mu\lp \{3\beta_1
\lp (\rhb_\mu^2- {1 \over a^2} \rp )^2
- 2 \frac{|\beta_2|}{a}  \, \rhb_\mu^3\rp \}
+\frac{3\beta_3}{a^2}\rhb_\mu^2.
\ee
This lagrangian density is stationary for
\be
\rhb^{(0)}_\mu = 0,\;\;
\rhb^{(\pm)}_\mu =
\frac{1}{4a\beta_1}
\lp [|\beta_2|\pm
\sqrt{\beta_2^2+8\beta_1\lp ( 2\beta_1-\beta_3\rp )}\rp ].
\ee
We have to chose the couplings so that $\rhb^{\lp (+\rp )}_\mu $
be a minimum and equal to $1/a$; this requirement gives
\be
|\beta_2|  = \beta_3,\;\; 4 \beta_1> \beta_3.
\ee
Under these conditions $\rhb^{\lp (0\rp )}_\mu $ is another
minimum, which we must require to lie higher than the
minimum at $\rhb^{\lp (+\rp )}_\mu $. This further strengthens the
above inequality to $3\beta_1 > \beta_3$.

The masses of the auxiliary fields turn out to be
\be
m^2_{\rho}=  { 6\over a^2}\lp(4\beta_1-\beta_3 \rp ),\;\;
m^2_{\phi} = {18\over a^2}\beta_3,
\;\;m^2_{\al}=  { 8\over a^2}\lp(2\beta_1+\beta_3 \rp ).
\ee
In conclusion the full classical lagrangian is
\be
{\cal{L}}_G={\cal{L}}_{YM}+{\cal{L}}_1+{\cal{L}}_2 +{\cal{L}}_3.
\ee

\section{Ward identities}

To determine the mass spectrum and identify the physical fields
we investigate the Ward identities.

We start with U(3) invariance and we assume that the color
symmetry is not
spontaneously broken. Therefore the effective action $\Gamma$
must be stationary
\be
\frac{\dep\Gamma}{\dep A^a_\mu (x)}  =
\frac{\dep\Gamma}{\dep\al^a_\mu (x)}=
\frac{\dep\Gamma}{\dep\ch_\mu (x)}=\frac{\dep\Gamma}{\dep\sg_\mu (x)}=0
\ee
for
\be
A^a_\mu (x)=\al^a_\mu (x)= 0, \;\;\;
\chi_{\mu}=\overline\chi_{\mu},\;\;\;
\sg_\mu (x)= \overline\sg_\mu.
\ee
Because of gauge invariance we have
\begin{eqnarray}
\delta\Gamma & = & \sum_{\mu,x} \lp [ \delta\ch_\mu (x)
\frac{\dep\Gamma}{\dep\ch_\mu (x)}
+\delta\sg_\mu (x)\frac{\dep\Gamma}{\dep\sg_\mu (x)} \rp ] \nonumber \\
  &   & +\sum_{a,\mu,x} \lp [ \delta A^a_\mu (x)
\frac{\dep\Gamma}{\dep A^a_\mu (x)}
+\delta\al^a_\mu (x)\frac{\dep\Gamma}{\dep\al^a_\mu (x)} \rp ] = 0 .
\end{eqnarray}

Introducing the explicit expressions for the variations and integrating
by parts we obtain
\begin{eqnarray}
\delta\Gamma & = & \sum_{\mu,x}\tht^a(x)
\lp\{\frac{2}{3}a\Delta^{(-)}_\mu
\lp [\al^a_\mu (x)\frac{\dep\Gamma}{\dep\ch_\mu (x)}
-A^a_\mu (x)\frac{\dep\Gamma}{\dep\sg_\mu (x)} \rp ] \rp . \nonumber \\
  &  & -\Delta^{(-)}_\mu\frac{\dep\Gamma}{\dep A^a_\mu (x)}
+2f^a_{bc}A^b_\mu (x)\frac{\dep\Gamma}{\dep A^c_\mu (x)}
+a\Delta^{(-)}_\mu \lp [ \sg_\mu (x)\frac{\dep\Gamma}{\dep A^a_\mu (x)}
\rp . \nonumber \\
  &   & \lp . -f^a_{bc}A^b_\mu (x)\frac{\dep\Gamma}{\dep A^c_\mu (x)}
+d^a_{bc}\al^b_\mu (x)\frac{\dep\Gamma}{\dep A^c_\mu (x)}\rp ]
+2f^a_{bc}\al^b_\mu (x)\frac{\dep\Gamma}{\dep\al^c_\mu (x)} \nonumber \\
  &   & \lp . +a\Delta^{(-)}_\mu \lp [- \ch_\mu (x)
\frac{\dep\Gamma}{\dep\al^a_\mu (x)} -f^a_{bc}\al^b_\mu (x)
\frac{\dep\Gamma}{\dep\al^c_\mu (x)}
-d^a_{bc}A^b_\mu (x)\frac{\dep\Gamma}{\dep\al^c_\mu (x)}\rp ]
\rp \} \nonumber\\
  &   & -\sum_{\mu,x}\tht^0(x)a\Delta^{(-)}_\mu
\lp\{\lp(1-a\sg_\mu (x)\rp)
\frac{\dep\Gamma}{\dep\ch_\mu (x)}+a\ch_\mu (x)
\frac{\dep\Gamma}{\dep\sg_\mu (x)}\rp .\nonumber \\
  &   & \lp . -\al^a_\mu (x)\frac{\dep\Gamma}{\dep A^a_\mu (x)}
+A^a_\mu (x)\frac{\dep\Gamma}{\dep\al^a_\mu (x)}
\rp \} = 0 .
\nonumber\\
\end{eqnarray}
We firstly assume $\theta_a =0 $. By taking the derivative wr to
$\chi_{\nu}$ and to $ \sigma_{\nu}$ we get at the minimum
\begin{eqnarray}
\lp (1-a\sgb_\mu\rp )\frac{\dep^2\Gamma}{\dep\ch_\nu (y)\dep\ch_\mu (x)}
+a\chb_\mu\frac{\dep^2\Gamma}{\dep\ch_\nu (y)\dep\sg_\mu (x)}=0
\nonumber \\
\lp (1-a\sgb_\mu\rp )\frac{\dep^2\Gamma}{\dep\sg_\nu (y)\dep\ch_\mu (x)}
+a\chb_\mu\frac{\dep^2\Gamma}{\dep\sg_\nu (y)\dep\sg_\mu (x)}=0 .
\end{eqnarray}
Analogously we assume $\theta^0=0$ and take the derivatives with respect
to $A_{\mu},\al_{\mu}$
\begin{eqnarray}
\lp (1-a\sgb_\mu\rp )\frac{\dep^2\Gamma}{\dep A_\nu (y)\dep A_\mu (x)}
+a\chb_\mu\frac{\dep^2\Gamma}{\dep A_\nu (y)\dep\al_\mu (x)}=0
\nonumber \\
\lp (1-a\sgb_\mu\rp )\frac{\dep^2\Gamma}{\dep\al_\nu (y)\dep A_\mu (x)}
+a\chb_\mu\frac{\dep^2\Gamma}{\dep\al_\nu (y)\dep\al_\mu (x)}=0 .
\label{Ward}
\end{eqnarray}
These equations show that in general there is a combination of the
fields $\chi_{\mu},\sg_{\mu}$
\be
\ch '_\mu(x)  =  \frac{1}{a}\lp [-s_\mu\lp (1-a\sg_\mu (x)\rp )
+a c_\mu\ch_\mu (x)\rp ]
\ee
and a combination of the fields $A_{\mu}, \al_{\mu}$
\be
A '_\mu(x)  =  -s_\mu \al_\mu (x)
+c_\mu A_\mu (x),
 \nonumber \\
\ee
with
\be
c_\mu=\frac{1-a\sgb_\mu}
{\lp[\lp (1-a\sgb_\mu \rp )^2+a^2\chb_\mu^2\rp ]^{\frac{1}{2}}},\;\;\;\;
s_\mu=\frac{a\chb_\mu}
{\lp[\lp (1-a\sgb_\mu \rp )^2+a^2\chb_\mu^2\rp ]^{\frac{1}{2}}},
\ee
which are massless. These are the physical fields. The actual
auxiliary fields are the orthogonal combinations

\begin{eqnarray}
\sg '_\mu(x) & = & \frac{1}{a}\lp [1-c_\mu\lp (1-a\sg_\mu (x)\rp )
-a s_\mu\ch_\mu (x)\rp ]\nonumber \\
\al '_\mu(x) & = & c_\mu \al_\mu (x)
+s_\mu A_\mu (x) .
\end{eqnarray}
The rotation to the primed fields is obtained by multiplying
$D_{\mu}$ by an element of the center of SU(3).

In SU(3) invariant theories we have only Eq.~(\ref{Ward}), so that
a mass for both abelian fields is no longer forbidden. In this
case both abelian fields are auxiliary.

\section{The background gauge and the BRS symmetry}

In the present regularization a background field can be introduced
in close analogy with the continuum (see for example
\cite{Abb}~and references therein) by performing a shift of the
gauge fields. We define a background covariant derivative,
which depends solely on the background fields, and the quantum
fluctuaction wrt these fields
\be
D_\mu(x)=D_{B,\mu}(x)+Q_\mu(x)
\ee
where
\begin{eqnarray}
D_{B,\mu}(x) & = & \lp[\frac{1}{a}-\sg_{B,\mu}(x)+i\ch_{B,\mu}(x)\rp]1\!\!1
+\lp [iA^a_{B,\mu}(x)-\al^a_{B,\mu}(x) \rp ]T_a, \nonumber\\
Q_{\mu}(x) & = & \lp[-\sg_{Q,\mu}(x)+i\ch_{Q,\mu}(x)\rp]1\!\!1
+\lp [iA^a_{Q,\mu}(x)-\al^a_{Q,\mu}(x) \rp ]T_a.
\end{eqnarray}
A gauge transformation of the covariant derivative $D_\mu$ 
\be
D'_\mu(x)=\lp [D_{B,\mu}(x)+Q_{\mu}(x)  \rp ]'=
g(x)\lp [D_{B,\mu}(x)+Q_{\mu}(x)  \rp ]g^{\dagger} (x+\mu).
\ee
can be interpreted, among the others, in the two following ways
{\it I interpretation}
\begin{eqnarray}
\lp (D_{B,\mu}(x) \rp )' & = & D_{B,\mu}(x),\nonumber\\
\lp (Q_{\mu}(x)  \rp )' & = &
g(x)\lp [D_{B,\mu}(x)+Q_{\mu}(x)  \rp ]g^{\dagger} (x+\mu)
-D_{B,\mu}(x)
\end{eqnarray}
{\it II interpretation}
\begin{eqnarray}
\lp (D_{B,\mu}(x) \rp )' & = & g(x)D_{B,\mu}(x)g^{\dagger} (x+\mu),
\nonumber\\
\lp (Q_{\mu}(x)  \rp )' & = &
g(x)Q_{\mu}(x) g^{\dagger} (x+\mu)
\end{eqnarray}
According to the first interpretation the background derivative
is invariant, while following the second interpretation it
transforms as the full covariant derivative. In the second case
the quantum fluctuaction undergoes a rotation like a matter field
in the adjoint representation.

The presence of the
background field enables us to introduce a gauge fixing term which
breaks the symmetry wrt the first interpretation, while preserving
the symmetry according to the second one.
So doing we shall obtain an effective action which is a gauge invariant
functional of the background field.

To define the gauge fixed theory we follow, for example,
ref.~\cite{vaBRS}, therefore we build a quantum lagrangian
renormalizable by power counting, BRS invariant and with zero
ghost number. The fundamental fields of the quantum theory are
\be
D_{B,\mu},Q_{\mu}(x),c(x),\cb(x),b(x)
\ee
where $c(x),\,\cb(x)$ are scalar Grassmann fields with,
respectively, positive and negative unit ghost number and canonical
dimension equal to 1 while $b (x)$ is a scalar c-number field
with vanishing ghost number and canonical dimension equal to 2;
the gauge quantum and background
fields obviously have vanishing ghost number.

We now determine the
equations for a BRS transformation of the various fields.
It is worthwhile noticing that the BRS symmetry corresponds to the
gauge symmetry broken by the gauge fixing term, therefore we determine
the BRS equations starting from those for an infinitesimal gauge
transformation according to the first interpretation which are,
for $ g(x)\simeq 1\!\!1 -i\tht^a (x) T_a$
\begin{eqnarray}
\delta D_{B,\mu}(x) & = & 0 \nonumber \\
\delta Q_\mu (x) & = & -i \tht^a (x) T_a D_\mu (x) +i D_\mu (x)
\tht^a (x+\mu) T_a
\end{eqnarray}

A BRS transformation is obtained by means of the~$s$ operator, whose
action on the various fields is specified by the following equations
\begin{eqnarray}
s\,D_{B,\mu} & = & 0,\nonumber \\
s\,Q_{\mu}(x) & = &  -i c(x) Q_\mu(x)+iQ_\mu c(x+\mu)\nonumber \\
  &   & -i c(x) D_{B,\mu}(x)+iD_{B,\mu} c(x+\mu),\nonumber \\
s\,c(x) & = & -iK(x), \nonumber \\
s\,\cb(x) & = & b(x), \nonumber \\
s\,b(x) & = & 0,
\end{eqnarray}
and the quantity~$K(x)$ is determined so as to obtain the
nilpotency of the~$s$ operator, namely
\be
K(x)=c(x)c(x).
\ee

The quantum theory is defined by the path integral
\be
Z\lp [ D_{B,\mu}(x)\rp] = \int {\cal{D}}Q_\mu (x){\cal{D}}c(x)
{\cal{D}}\cb(x){\cal{D}}b(x)\exp\lp\{-\sum_{x,\mu}
\lp[{\cal{L}}_G(x)+{\cal{L}}_{BRS}(x)\rp ]\rp \}
\ee
where
\begin{eqnarray}
{\cal{L}}_{BRS}(x) & = &- \lambda \beta \mbox{Tr}\,s\,
\lp \{\cb (x) \lp[{\cal{G}}(x)-b(x)\rp]\rp\} \nonumber\\
   & = & - \lambda \beta \mbox{Tr}\,\lp\{b(x){\cal{G}}(x)-b(x)b(x)\rp\}
+ \lambda \beta \mbox{Tr}\,\lp\{\cb(x)\,s\,{\cal{G}}(x)\rp\}\nonumber\\
   & = & {\cal{L}}_{gf}(x)+{\cal{L}}_{ghost}(x).
\end{eqnarray}
The quantity ${\cal{G}}(x)={i\cal{G}}^0(x)1\!\!1+{\cal{G}}^a(x)T_a$
is the gauge fixing constraint and $\lambda$ is a real positive parameter.
We can get rid off the $b(x)$ field with a gaussian integration,
so obtaining
\be
{\cal{L}}_{gf}(x)=-\frac{ \lambda \beta}{2}{\cal{G}}^a(x){\cal{G}}^a(x).
\ee
A gauge fixing term which preserves the exact gauge symmetry for
transformations of the background field is
\begin{eqnarray}
{\cal{G}}(x) & = & -i\sum_{\mu}\lp[ D_{B,\mu}^{\dag}\lp(x-{\mu}\rp)
Q_{\mu}\lp(x-{\mu}\rp)-Q_{\mu}(x)D_{B,\mu}^{\dag}(x)\rp ].
\end{eqnarray}
Following the second interpretation ${\cal{G}}(x)$ varies according to
\be
\lp({\cal{G}}(x)\rp)'  =  g(x){\cal{G}}(x)g^{\dag}(x).
\ee
As a consequence the gauge fixing term is invariant under gauge transformations
of the background field and the effective action is a gauge invariant
functional of the latter.

\section{Summary}

We have reconsidered a lattice regularization of gauge 
theories which makes use of auxiliary fields in order to enforce exact 
gauge invariance with noncompact fields. The form of the covariant 
derivative, for n $>$ 2, is the same for U(n) and SU(n) theories.
This means that the physical abelian field of the U(n) theory must 
become an additional auxiliary field in the SU(n) theory. 
This can be guaranteed at the quantum level by breaking explicitely 
the U(n) symmetry in such a way as to generate a divergent mass for 
this field. The terms of the lagrangian which realize this condition
have been exhibited and their effect investigated. The regularization 
can now be used on essentially the same footing for every n.

We have also investigated the Ward identities of the effective 
action, confirming that the mass spectrum has the desired properties. 
Finally  we have formulated the theory in the background gauge and 
written the BRS identities, showing that a perturbative treatment 
can be done in close analogy with the continuum,
avoiding the cumbersome expansion of the link variables.

\section{Appendix A}

In this Appendix we report the explicit expression of $ {\cal{L}}_2  $
\begin{eqnarray}
& & {\cal{L}}_2  =  \beta_2 { 1\over a} \sum_\mu
\lp [ - det\,D_\mu(x) - det\,D_\mu^{\dagger}(x) + \frac{2}{a^3} \rp]
\nonumber \\
  & = & \beta_2 { 1\over a}\sum_\mu \lp \{ \frac{6}{a^2}\sg_\mu(x)
-{6\over a} \lp (\sg^2_\mu(x)-\ch^2_\mu(x)\rp)
- {2\over a}  \lp ( A ^a_\mu(x)A ^a_\mu(x)
-\al ^a_\mu(x)\al ^a_\mu(x) \rp ) \rp . \nonumber \\
  &   & +2\sg^3_\mu (x)  -6\sg_\mu(x)\ch^2_\mu(x)
+4\ch_\mu(x)A^a_\mu(x)\al^a_\mu(x)  \nonumber \\
  &   &  +2\sg_\mu(x) \lp(
A^a_\mu(x)A^a_\mu(x)-\al^a_\mu(x)\al^a_\mu(x)\rp )
\nonumber \\
  &   & -4\sum_{a=1}^8 d^8_{aa}\lp [2A^8_\mu(x)A^a_\mu(x)\al^a_\mu(x)
+\al^8_\mu(x)\lp (A^a_\mu(x)A^a_\mu(x)-\al^a_\mu(x)\al^a_\mu(x)\rp )\rp ]
\nonumber \\
  &   & -4\sum_{a=4}^7 d^3_{aa}\lp [2A^3_\mu(x)A^a_\mu(x)\al^a_\mu(x)
+\al^3_\mu(x)\lp (A^a_\mu(x)A^a_\mu(x)-\al^a_\mu(x)\al^a_\mu(x)\rp )\rp ]
\nonumber \\
  &   & -8 d^1_{57}\lp [A^1_\mu(x)\lp (A^5_\mu(x)\al^7_\mu(x)
+A^7_\mu(x)\al^5_\mu(x) \rp ) +\al^1_\mu(x)\lp (A^5_\mu(x)A^7_\mu(x)
+\al^5_\mu(x)\al^7_\mu(x) \rp )\rp ]
\nonumber \\
  &   & -8 d^1_{46}\lp [A^1_\mu(x)\lp (A^4_\mu(x)\al^6_\mu(x)
+A^6_\mu(x)\al^4_\mu(x) \rp ) +\al^1_\mu(x)\lp (A^4_\mu(x)A^6_\mu(x)
+\al^4_\mu(x)\al^6_\mu(x) \rp )\rp ]
\nonumber \\
  &   & -8 d^2_{47}\lp [A^2_\mu(x)\lp (A^4_\mu(x)\al^7_\mu(x)
+A^4_\mu(x)\al^5_\mu(x) \rp ) +\al^2_\mu(x)\lp (A^4_\mu(x)A^7_\mu(x)
+\al^4_\mu(x)\al^7_\mu(x) \rp )\rp ]
\nonumber \\
  &   & \lp .-8 d^2_{56}\lp [A^2_\mu(x)\lp (A^5_\mu(x)\al^6_\mu(x)
+A^6_\mu(x)\al^5_\mu(x) \rp ) +\al^2_\mu(x)\lp (A^5_\mu(x)A^6_\mu(x)
+\al^5_\mu(x)\al^6_\mu(x) \rp )\rp ]\rp\}
\nonumber \\
\end{eqnarray}

\end{document}